\documentclass[mathleft
]{an}
\usepackage{graphicx}
\usepackage{times}
\begin{document}

\Pagespan{1}{}
\Yearpublication{2010}%
\Yearsubmission{2010}%
\Month{11}%
\Volume{999}%
\Issue{88}%

\newcommand{\Deg}{\hbox{${}^\circ$}}
\newcommand{\Min}{\hbox{${}^{\prime}$}}
\newcommand{\Sec}{\hbox{${}^{\prime\prime}$}}
\newcommand{\minp}{\hbox{${}^{\prime}$\llap{.}}}

\title{A search for radio counterparts to $Chandra$ ULX candidates}

\author{D. P\'erez-Ram\'{\i}rez\inst{1}\fnmsep\thanks{D. P\'erez-Ram\'{\i}rez: \email{dperez@ujaen.es}\newline}
\and  M. Mezcua\inst{2}
\and  S. Leon\inst{3}
\and  M.D. Caballero-Garc\'{\i}a\inst{4}
}
\titlerunning{A search for radio counterparts to $Chandra$ ULX candidates}
\authorrunning{D. P\'erez-Ram\'{\i}rez et al.}
\institute{
Universidad de Ja\'en, Campus Las Lagunillas, s/n, E-23070, Ja\'en, Spain
\and
Max Planck Institute for Radio Astronomy, Auf dem H\"ugel 69, 53121 Bonn, Germany
\and
Joint Alma Observatory/ESO, Las Condes, Santiago, Chile
\and 
University of Cambridge, Institute of Astronomy, Cambridge CB3 0HA, UK                     
}

\received{}
\accepted{}
\publonline{later}


\abstract{%
We present a systematic search for  radio counterparts to Ultra
Luminous X-ray (ULX) source candidates based on a cross-correlation of
the Swartz et al. (2004) ULX catalogue based on $Chandra$ data and the
FIRST radio survey.  We find seven cases of conspiscuous peaks of
radio emission that could be associated to ULX sources. Among these
seven ULX radio candidates, three X-ray sources are located within 5''
of the FIRST radio peaks.  These three cases are shown and discussed
individually.}

\maketitle

\section{Introduction}

The nature of Ultraluminous X-ray Sources (ULXs) is still under
debate. Different models have been proposed to explain the possible
physical mechanisms at work that give rise to the high X-ray
luminosities observed (e.g. extragalactic microquasars, King et
al. 2001; stellar mass black holes with super Eddington radiation,
 Begelman 2002;background AGN along the line of sight, Irwin, Bregman
\& Athey 2004; intermediate mass black holes, Colbert \& Mushotzky
1999; or a consequence of a new ultraluminous accretion state, i.e. a
transition between Eddington and super-Eddington accretion flows,
Gladstone, Roberts \& Done 2009).

New X-ray surveys and associated ULX catalogues now available have
compiled lists of potential ULX candidates. These catalogues are based
on data from specific missions extracted from their corresponding
archives. Earlier catalogues were based on ROSAT HRI data, as the one
by Colbert \& Ptak (2002) that registered a total of 54 galaxies with
87 possible associated ULX candidates. These numbers increased with
the catalogue by Liu \& Mirabel (2005), who compiled individual cases
from published literature until April 2004 resulting in a total of 85
galaxies and 229 ULX new cases. Liu \& Bregman (2005) presented a more
complete survey based also on ROSAT HRI data, where the number of
recorded galaxies in the catalogue reached 313, and their possible
associated ULXs increased to 562. In a previous work (see
S\'anchez-Sutil et al. 2006) we initiated a systematic search of ULX
radio counterparts mainly based on the cross-identification of the Liu
\& Bregman (2005) catalogue and the FIRST catalogue, resulting in a
total of 11 previously unreported matches, with a estimated positional
uncertainty of $\sim$ 10$\arcsec$.

The most promising data at the present time come from the $Chandra$
mission due to its high spatial resolution and improved astrometry. In
this sense, the catalogue by Swartz et al. (2004), hereafter Sw04,
based on $Chandra$ data is the only compiled list of potential ULXs
with an estimated positional uncertainty of about 1-2$\arcsec$.

The study of ULX radio counterparts can give us a clue to establish
the ULX morphology, spectral index and variability properties.
However, a radio counterpart for an ULX has only been found in a very
few cases (Kaaret et al. 2003; K\"ording et al. 2005; Miller et
al. 2005; S\'anchez-Sutil et al. 2006).

We continue searching for possible radio counterparts to ULX
candidates, now taking advantage of the published $Chandra$ data
catalogue. The work has been done following a similar approach to the
one cited in Mushotzky (2004). The main results of our work reveal
several new positional coincidences of FIRST radio sources with
potential ULXs. In this paper we include both radio maps and
individual discussion of the coincidence cases of special interest.

\section{Cross-identification and procedure}

\begin{table*}
\caption[]{\label{coin} Selected ULX sources with possible radio counterparts}
\begin{tabular}{ccccccccc}
\hline
\hline
NGC    &  ULX  &  $\alpha_{{\rm J2000.0}}$  & $\delta_{{\rm J2000.0}}$ & Separation & Radio vs.  &   Peak
         &Integrated &  RMS \\
       &  Name &         (FIRST)            &        (FIRST)           &  from nucleus &X-ray  offset    &  Flux 
Density  & Flux       & \\
       &       &                            &                          & (\Min) & (\Sec)       &    (mJ
y/beam)       &  (mJy)    & (mJy/beam)     \\
\hline

4490  & X1 & $12^{\rm h}$ $30^{\rm m}$ 29\fs46  & $+41$\Deg~39\Min~27.4\Sec & 1.5 & 1.1  &    3.34 & 
 7.17  &  0.156\\

5194  & X2 & $13^{\rm h}$ $29^{\rm m}$ 50\fs65  & $+47$\Deg~11\Min~54.6\Sec & 0.4 & 0.4  &    1.20 & 
 3.20  &  0.139\\

5775  & X4 & $14^{\rm h}$ $54^{\rm m}$ 00\fs98  & $+03$\Deg~31\Min~29.0\Sec & 1.4 & 4.1 &    2.94 & 
 4.04  &  0.139\\

\hline
\end{tabular}
\end{table*}

We cross-identified the Sw04 catalogue, based on $Chandra$ data, and
the FIRST radio catalogue.  Sw04 compiled a sample of 154 ULX
candidates identified in 82 galaxies. These galaxies belong to
different Hubble morphological types and also present distinct
characteristics which make the sample more complete. The catalogue
sample was also cleared for ULXs with positions located at less than
5$\arcsec$ from the nucleus.

We used the catalogue of the VLA FIRST survey (Becker et al. 1995)
that was released on April 2003 and was derived from the 1993 through
2002 observations (White et al. 1997).  This survey presents a good
sensitivity (rms $\sim$ 0.15 mJy) and angular resolution ($\sim
5^{\prime\prime}$).

We based the cross-identification on the positional coincidence of the
FIRST and the $Chandra$ ULX entries within less than a certain error
which depends mainly on the relative accuracy of the instruments
involved. We established a radius of $5^{\prime\prime}$, which is
roughly the maximum offset typically expected from the combined
FIRST-$Chandra$ astrometric errors.  Not all the Sw04 sample galaxies
were observed by the FIRST survey.  We found FIRST radio data for 37
out of the 82 galaxies from the Sw04 sample, so only 43$\%$ of the
Sw04 ULX parent galaxies were present.

From these 37 FIRST galaxies, half of them presented their ULX
candidates located in regions where no radio emission was detected in
the radio maps. The other half of the galaxy sample, this is 8 out of
19, presented very poor signal to noise ratio radio maps.  It was thus
not possible to check radio coincidences for these cases.

Although the Sw04 catalogue does not contain any candidate within 5''
of the nucleus, in 4 out of the 11 galaxies left, the matched radio
source lies closer to the nucleus than the ULX candidate which might
imply that the source could be nuclear in origin, and therefore, not a
true ULX.

We ended up with 7 cases. We visually checked the matches and looked
for compact radio counterpart candidates, which are the expected
appearance of ULXs at the limited FIRST angular resolution and are
compatible with the most plausible physical scenarios. However,
extended radio sources were also considered if they were located
within knotty optical features or spiral arms as possibly related to
HII regions, since HII regions represent another suggested ULX
scenario (Miller et al. 2005). Among these matches we report in the
current work three cases for which the offset between the X-ray source
and the peak of compact radio emission is less than 5'', and are
located at distances larger than 20'' from the nucleus within the
galaxy disk. We present details for these cases in Table~1, where the
FIRST estimated coordinates, the separation of the ULX with respect to
the nucleus of the galaxy as given in NED, the radio/X-ray offset,
peak flux density, integrated flux density, and rms are given
respectively from Columns~3 to 9.  The numeration used to coin the
ULXs responds to the order of these sources as they appear in the Sw04
list.

We looked for possible counterparts in the SIMBAD database for each of
the selected sources, finding only a previous citation by Roberts et
al. (2002) for one of the ULX candidates in the galaxy NGC 4490, who
suggested that the source could be a supernova remnant (SNR).
Figures~1 to 3 show the optical and radio fields to illustrate their
positional coincidence. The optical images were retrieved from the
Digitized Sky Survey (DSS) (Lasker et al. 1990), while the radio ones
were taken from FIRST.

\section{Comments on selected cases}

This section is devoted to provide a more detailed individual
discussion on the most interesting cases found in the
cross-identification process.

\subsection{ULX~1 in NGC~4490}

   \begin{figure*}
   \centering
\resizebox{\hsize}{!}{\includegraphics[scale=0.3]{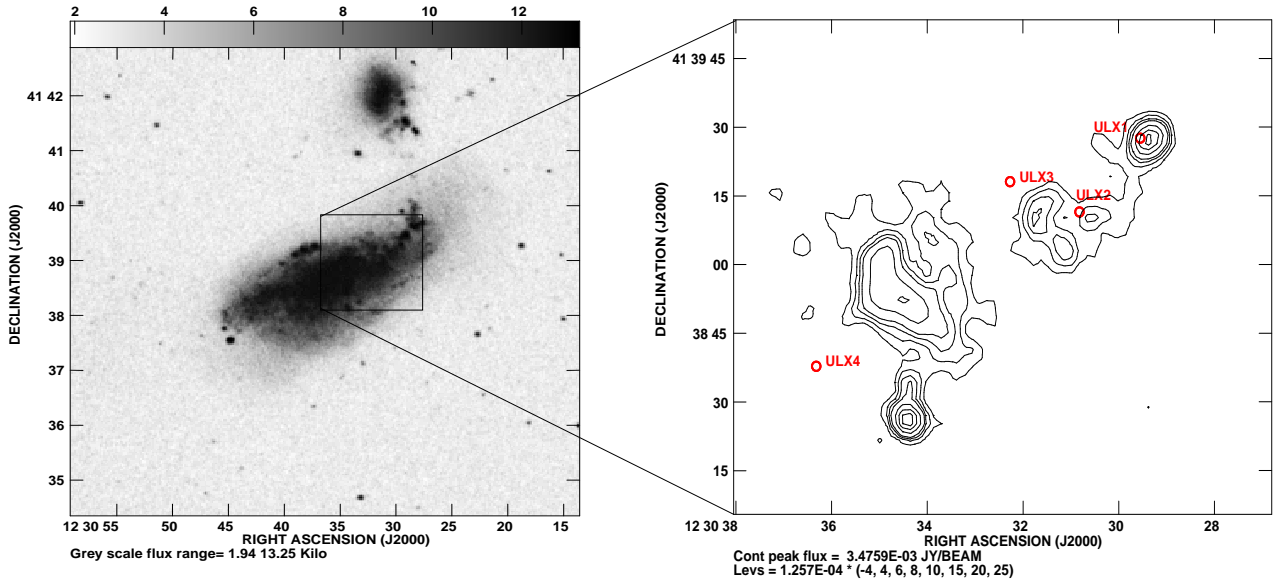}}
      \caption{Left panel: Optical DSS image for NGC~4490. Right panel: 
FIRST radio image contoured in units of the rms noise ($\sim$0.15
mJy/beam). These values also apply to the rest of figures. The
locations of all ULX candidates listed in the Sw04 catalogue for this
galaxy are shown. ULX~1 is located within the disk of NGC~4490.}
      \label{ngc4490}
   \end{figure*}

NGC 4490 is a galaxy classified as SBm, a system close to be
edge-on. It presents a knotty center but does not exhibit apparent
nucleus or bulge.

Sw04 listed 5 potential ULXs associated to this galaxy. After close
inspection we found that 4 of them are within the galaxy disk. The
FIRST radio emission is limited to a smaller region close to the
nucleus rather than covering the whole disk.  We found a clear
coincidence of ULX~1 with one of the peaks detected from the FIRST
observations, and a second one, ULX~2, possibly associated to another
emission peak (see Fig.~1). Details on the peak flux and the FIRST
location of the possible counterpart to ULX~1 are given in
Table~1. Roberts et al. (2002) identified this source as CXOU
J123029.5+413927. After the analysis of its spectral properties they
suggested that the source could be a SNR. However, in a following work
by Gladstone \& Roberts (2009), it was found that both flux and
spectrum for the X-ray source changed, arguing against the previously
suggested origin, and indicating that could be an X-ray binary
instead.

\subsection{ULX~2 in NGC~5194}

   \begin{figure*}
   \centering
\resizebox{\hsize}{!}{\includegraphics[scale=0.3]{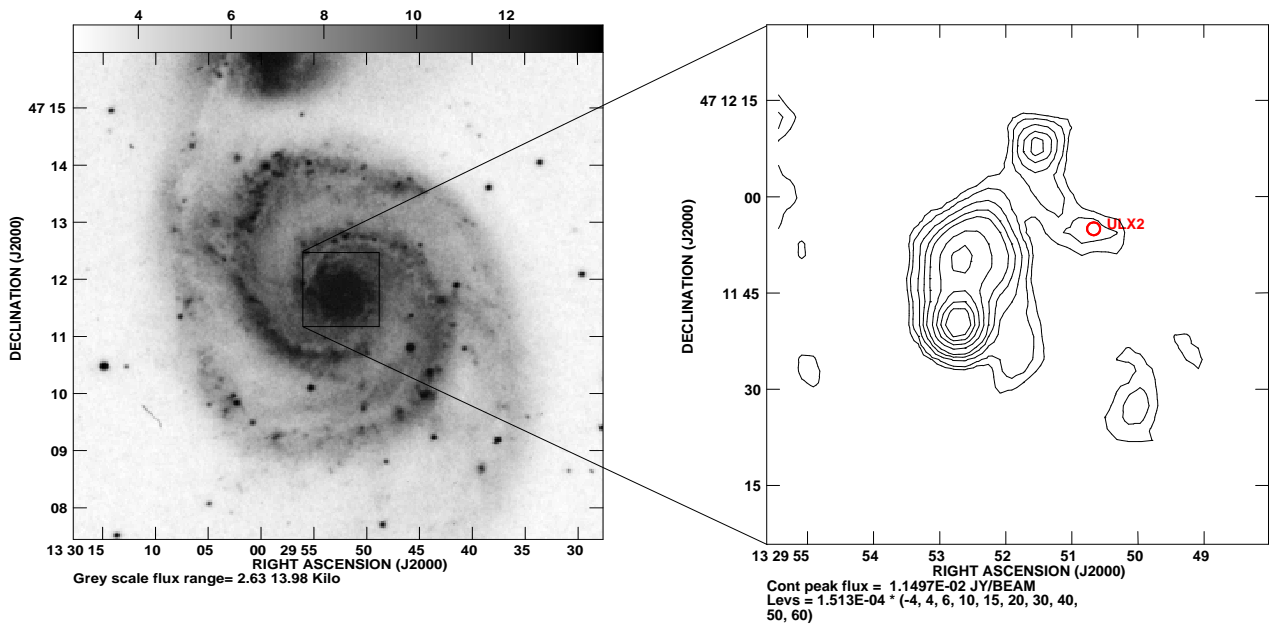}}
      \caption{As in Fig~1, now for NGC 5194. ULX~2 is located within 
the disk of NGC~5194.}
      \label{ngc5194}
   \end{figure*}

NGC 5194 is an almost face-on grand-design spiral galaxy with
bisymmetric spiral arms of molecular gas. It is a low-luminosity
starburst galaxy with faint non-thermal radio core.  It is also known
as the Whirlpool Nebula and it forms an interacting system with its
companion NCG 5195 (Keel et al. 1985). 

The radio emission of this galaxy is not concentrated in the nucleus
as in the previous case but it is extended over the galaxy disk.  Sw04
listed a total of 7 potential ULXs for this galaxy. We found an
evident coincidence of ULX~6 with the nucleus of the companion galaxy
NGC~5195. For the rest of ULXs, except for ULX~2, there was no
correspondence at all between the FIRST contours and the locations of
the $Chandra$ sources. In the case of ULX~2, it is located almost
coincident with a peak of radio emission, with a separation of 0.4''
(see Figure~2).  Details of the peak flux estimations are given in
Table~1 .

This galaxy was also studied in the work by S\'anchez-Sutil et
al. (2006). Three matches were found after cross-identification, but
they were discarded as possible nuclear missidentifications. However,
in the present work, we found that the position of ULX~2 from the
FIRST contour map cannot be considered a nuclear coincidence.

\subsection{ULX~4 in NGC~5775}

\begin{figure*}
   \centering
\resizebox{\hsize}{!}{\includegraphics[scale=0.3]{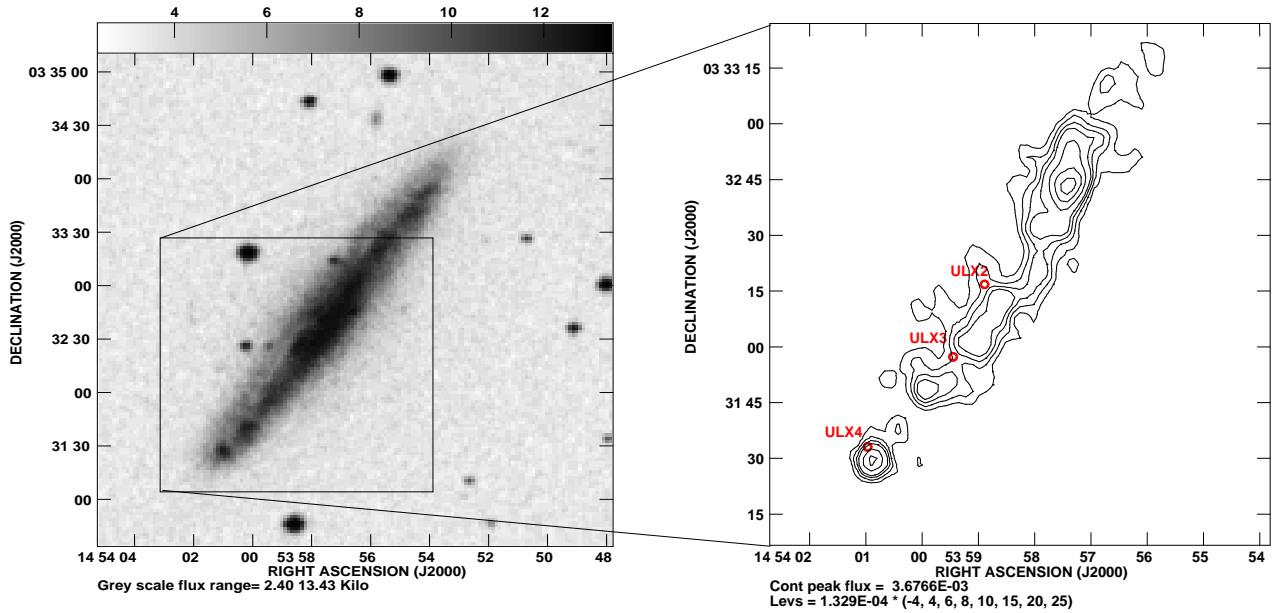}}
      \caption{As in Fig~1, now for NGC~5775. ULX~4 is located within the
disk of NGC~5775.}
      \label{ngc5775}
   \end{figure*}

NGC~5775 was found to be an interacting galaxy with its neighbour
face-on spiral galaxy NGC~5774. Estimations based on FIR ratio by
Rossa \& Dettmar (2003) suggested this galaxy to be a starburst-type
galaxy.

From H$\alpha$ images a possible tidal interaction has been suggested
to happen, which together with the numerous compact sources at the
south-east, presumably HII regions, could be an indication that star
formation may be occurring well above the disk (Collins et al. 2000).

Sw04 give 4 possible ULXs associated to this galaxy, none of them
coincident with the location of the nucleus given by NED. Three
$Chandra$ ULXs are located within the galactic disk in the outskirts
of the contour emission map (see Figure~3). There is only one clear
association of a FIRST radio emission peak for the location of
ULX~4. This source is coincident with the source reported (numbered
47) in Ghosh et al. (2009). Detailed estimations for the peak and the
FIRST location of the source are given in Table~1 .

\section{Conclusions}

We present results of a cross-identification of the Sw04 ULX catalogue
based on $Chandra$ data and the FIRST radio catalogue. We found FIRST
radio data for 37 out of the 82 galaxies from the Sw04 sample. We
ended up with 7 cases. We show in this paper three cases for which the
X-ray sources lie within 5'' of the compact radio peak locations.

We did not find evidence of background QSO/AGN association based on
location, as they are within the galaxy disks.  None of them can be
neither associated with any obvious clump of star formation visible in
the optical image. This possibility needs to be confirmed with new
observational data and H$\alpha$ images of the host galaxies.

Further higher-angular resolution radio observations and improved
$Chandra$ astrometry of the data are in progress. A new collection of
new ULX candidates associated to more recent observed galaxies by
$Chandra$ is being compiled which will make up an extended version of
the ULXs catalogue used for this work. Investigation of new
cross-correlated cases will make possible to identify new cases and
establish the true nature of the radio emission associated to these
objects.

\acknowledgements

We thank the referee for constructive comments to improve the clarity
of the paper. The research of DPR has been supported by the
Universidad de Ja\'en (Spain). DPR acknowledges A. Castro-Tirado for useful 
discussions, J. Mart\'{i} for suggesting the study of this set of galaxies, 
and finally, A. Mu\~{n}oz and J.R. S\'anchez for their cooperation in 
making the figures. This research made use of the SIMBAD database, operated 
at the CDS, Strasbourg, France.

\appendix


\begin{thebibliography}{}


\bibitem[Becker et al. (1995)]{×}
Becker, R.~H., White, R.~L., \& Helfand, D.~J.\ 1995, ApJ, 450, 559 

\bibitem[Begelman (2002)]{beg02}
Begelman, M., 2002, ApJ, 568, L97

\bibitem[Colbert \& Mushotzky (1999)]{cp1999}
Colbert, E. J. M., Mushotzky, R. F., 1999, ApJ, 519, 89

\bibitem[Colbert \& Ptak (2002)]{cp2002}
Colbert, E. J. M., Ptak, A. F., 2002, ApJS, 143, 25

\bibitem[Gladstone, Roberts \& Done (2009)]{grd09}
Gladstone, J.C., Roberts, T.P. and Done, C., 2009, MNRAS, 297, 1836

\bibitem[Gladstone \&  Roberts(2009)]{gr09}
Gladstone, J.C., and Roberts, T.P., 2009, MNRAS, 397, 124

\bibitem[Ghosh et al. (2009)]{ghosh09}
Ghosh, K. K., Saripalli, L., Gandhi, P., Foellmi, C., Guti\'errez, C.M. and L\'opez-Corredoira, M., 2009, AJ, 137, 3263

\bibitem[Kaaret et al. (2003)]{k03}
Kaaret, P., Corbel, S., Prestwich, A. H., Zezas, A., 2003, Science, 299, 365

\bibitem[Keel et al. (1985)]{ke85}
Keel, W.C., Kennicutt, R.C., Jr., Hummel, E. et al., 1985, AJ, 90,708

\bibitem[King et al. (2001)]{king01} 
King, A.R., Davies, M.B., Ward, M.J., Fabbiano, G., Elvis, M.,2001, ApJ, 
552, L109

\bibitem[K\"ording et al. (2005)]{k2005}
K\"ording, E., Colbert, E., Falcke, H., 2005, A\&A, 436,427

\bibitem[Lasker et al. (1990)]{dss}
Lasker, et al., 1990, AJ, 99, 2019

\bibitem[Liu \& Bregman (2005)]{lb2005}
Liu, J.-F., Bregman, J. N., 2005, ApJS, 175, 59

\bibitem[Liu \& Mirabel (2005)]{lm2005}
Liu, Q. Z., Mirabel, I. F., 2005, A\&A, 429, 1125

\bibitem[Miller et al. (2005)]{mil2005}
Miller, N.A., Mushotzky, R.F., Neff, S.G., 2005, ApJ, 623, L109

\bibitem[Mushotzky (2004)]{mus04}
Mushotzky, R., 2004, Prog. Theor. Phys. Supp., 155, 27

\bibitem[Roberts et al. (2002)]{rob02} 
Roberts, T.P., Warwick, R.S., Ward, M.J. et al., 2002, MNRAS, 337, 677

\bibitem[S\'anchez-Sutil et al. (2006)]{ss06} 
S\'anchez-Sutil, J.R., Mun\~{n}oz-Arjonilla, A.J., Mart\'{i}, J.,
Garrido, J.L., P\'erez-Ram\'{i}rez, D., Luque-Escamilla, P., 2006,
A\&A, 452, 739

\bibitem[Swartz et al. (2004)]{sw04}
Swartz, D., Ghosh, K.K., Tennant, A.F., et al.,  2004, ApJS, 154, 519

\bibitem[White et al. (1997)]{w1997}
White, R. L., Becker, R. H., Helfand, D. J., Gregg, M. D., 1997, ApJ,
475, 479

\end{thebibliography}
\end{document}